\documentclass[twocolumn,aps,psfig,epsfig,bm,showpacs,superscriptaddress]{revtex4} 
\usepackage{epsfig}
\usepackage{amssymb}

\begin{document} 

\newcommand{\vk}{{\vec k}} 
\newcommand{\vK}{{\vec K}}  
\newcommand{\vb}{{\vec b}}  
\newcommand{\vp}{{\vec p}}  
\newcommand{\vq}{{\vec q}}  
\newcommand{\vQ}{{\vec Q}} 
\newcommand{\vx}{{\vec x}} 
\newcommand{\vh}{{\hat{v}}} 
\newcommand{\tr}{{{\rm Tr}}}  
\newcommand{\be}{\begin{equation}} 
\newcommand{\ee}{\end{equation}}  
\newcommand{\half}{{\textstyle\frac{1}{2}}}  
\newcommand{\gton}{\stackrel{>}{\sim}} 
\newcommand{\lton}{\mathrel{\lower.9ex \hbox{$\stackrel{\displaystyle 
<}{\sim}$}}}  
\newcommand{\ben}{\begin{enumerate}}  
\newcommand{\een}{\end{enumerate}} 
\newcommand{\bit}{\begin{itemize}}  
\newcommand{\eit}{\end{itemize}} 
\newcommand{\bc}{\begin{center}}  
\newcommand{\ec}{\end{center}} 
\newcommand{\bea}{\begin{eqnarray}}  
\newcommand{\eea}{\end{eqnarray}}

\title{Differential freezeout and pion interferometry at RHIC \\ from covariant transport theory}
 
\date{\today}
 
\author{D\'enes Moln\'ar}
\affiliation{Department of Physics, Ohio State University,
                174 West 18th Ave, Columbus, OH 43210}
\affiliation{Department of Physics, Columbia University, 
             538 West 120-th Street, New York, NY 10027}
\author{Miklos Gyulassy}
\affiliation{Department of Physics, Columbia University, 
             538 West 120-th Street, New York, NY 10027}

\begin{abstract} 
Puzzling discrepancies between recent 
pion interferometry data on $Au+Au$ reactions at 
$\sqrt{s}=130$ and $200$ $A$GeV 
from RHIC and predictions based on ideal hydrodynamics
are analyzed in terms of covariant parton transport theory. 
The discrepancies of out and longitudinal radii are significantly reduced
when the finite opacity of the gluon plasma is taken into account.
\end{abstract}

\pacs{12.38.Mh; 24.85.+p; 25.75.Gz; 25.75.-q}

\maketitle 

{\em Introduction.}
Decoupling, or freezeout of final state interactions,
 is a process with unique sensitivity
to the space time evolution of hadronic interactions.
It reflects the interplay between the decreasing opacity of the
system at late times and correlations induced
by collective expansion driven by high scattering rates  at early times.

In heavy-ion physics, information about the 
spacetime decoupling geometry can be obtained via
 identical boson (Hanbury-Brown and Twiss (HBT)) interferometry.
Recent two pion correlation data
in Au+Au at $\sqrt{s} = 130$ \cite{STARhbt,PHENIXhbt}
and $200$ $A$GeV/nucleon \cite{QM2002hbt} at RHIC
seem at first sight to indicate
a sudden freezeout that is difficult to reconcile
with the strong 
collective dynamics implied by the substantial elliptic flow
$v_2 \sim 0.1$ also observed\cite{STARv2,PHENIXv2}.

Ideal Euler hydrodynamics provides
one of the powerful covariant approaches 
to predict the collective flow pattern as well
as possible freezeout hypersurfaces
in heavy-ion collisions. 
However, this approach predicts an "out" radius, $R_{out}$ as defined below,
significantly larger than the "side" radius,
$R_{side}$\cite{DirkHBT,hydroHBT}.
On the other hand, it is possible that
the failure of non-dissipative hydrodynamics to
correctly describe the delicate space-time decoupling
geometry is 
due to the neglect of deviations from local equilibrium
throughout the evolution.
HBT predictions with hydrodynamics
are  based on an additional ad hoc (Cooper-Frye) postulate
that freezeout occurs on some  "thin" three-dimensional
hypersurface (typically, an isotherm). 
Only a detailed covariant transport theory can assess
the theoretical error introduced by this postulate. 

The simplest Lorentz covariant dynamical framework 
that can predict freezeout {\em self-consistently} 
is transport theory\cite{ZPC,nonequil,v2,finrange}.
In this approach the interaction rate is controlled by microscopic
differential cross sections, $d\sigma$.
As the system expands and rarefies, the scattering rate
decreases until the particles stop interacting.
In \cite{v2} it was emphasized that finite cross sections
are needed to account for the saturation of differential 
elliptic flow $v_2(p_\perp)$ observed 
for $p_\perp > 2$ GeV in Au+Au at RHIC\cite{STARv2,PHENIXv2}.

The influence of final state dissipation
on HBT was studied in a hybrid hydrodynamic/transport model in Ref. 
\cite{AdrianHBT}.
 In that work, hydrodynamical evolution was followed
only up to a hadronization isotherm $T(x^\mu)=T_c$. 
Subsequently, the decoupling of the hadron
gas was computed via the UrQMD hadronic transport model. 
However, the predicted $R_{out}/R_{side} >1 $ increasing with
transverse momentum still fails to account for the observed decreasing
$R_{out}/R_{side} <1$. Similar hadronic transport results were reported in \cite{Humanic}.
This suggests that possible deviations from the local equilibrium assumption
prior to hadronization should also be explored.

Recently, a combined parton/hadron transport theory
approach was proposed  in \cite{Lin:2002gc}. The results suggest 
that the HBT radii are indeed sensitive to the parton cross section
during the dense partonic phase of the reaction. Unfortunately,
all the transport calculations above left open the delicate question
of Lorentz covariance of the numerical solutions.
In any case, we
note that no transport or hydrodynamic calculation has as yet
been able to reproduce the phenomenological decoupling source parametrizations 
fitted to that data  in \cite{Csorgo:2002ry}.

In this letter, we concentrate exclusively on the partonic transport
phase to isolate more clearly the influence of dissipative
partonic processes on the decoupling geometry and study in detail the question
of covariance. Preliminary results were
reported in \cite{Molnar:2002ax}.
We utilize the MPC numerical technique\cite{MPCsource},
and compute the covariant 
freezeout distributions 
for a wide range of RHIC initial conditions
 as a function of the partonic opacity.

{\em Two-particle HBT interferometry.}
For a chaotic boson freeze-out source $\rho(\vx, t)$,
the two-particle momentum correlation function $C(p_1, p_2)$ is 
given by the {\em space-time} Fourier transform of the source
\cite{MiklosHBT,Csorgo:1999sj,UrsHBT}. 
Although the relation cannot be inverted
(due to loss of phase information and 
the on mass-shell constraint), 
HBT measurements provide a unique test of 
freezeout distributions  and dynamical scenarios in heavy-ion collisions.

Conventionally,
the two-particle correlation function is expressed in terms of
the relative momentum 
$q^\mu \equiv p_1^\mu- p_2^\mu$ and average pair momentum
$K^\mu \equiv (p_1^\mu + p_2^\mu) / 2$.
The 'out-side-long' variables, $q_O$, $q_S$, and $q_L$,
are then defined via $\vq_{LCMS} \equiv (q_O, q_S, q_L)$
in the Longitudinal Center of Mass System (LCMS) reference frame where 
$K^\mu_{LCMS} = (\tilde K^0, K_\perp, 0, 0)$.
We denote here the LCMS spacetime coordinates by
$x^\mu_{LCMS} \equiv (\tilde t, x_O, x_S, x_L)$.

Experimentally,
the measured correlation function
(after correcting for Coulomb distortions)
is fitted with a Gaussian, which for central collisions and 
at midrapidity is constrained by symmetry to the form
\be
C(\vq, K) = 1+ \lambda(K) \exp\!\left[ -\sum_{i=O,S,L} q_i^2 R_i^2(K)\right] \ .
\ee
Here $R_O$, $R_S$ and $R_L$ are the 'out', 'side', and 'long' HBT radii.
For a {\em perfectly Gaussian} source,
the correlation function is Gaussian and \cite{UrsHBT}
\bea
R_O^2(K) &=& \langle \Delta x_O^2 \rangle_K
        + v_\perp^2 \langle \Delta \tilde t^2 \rangle_K
        - 2 v_\perp  \langle \Delta x_O \Delta \tilde t \rangle_K
\nonumber \\
R_S^2(K) &=& \langle \Delta x_S^2 \rangle_K, \qquad{\rm and} \quad R_L^2(K) = \langle \Delta x_L^2\rangle_K \ , \ \ 
\label{HBTradii}
\eea
where $v_\perp \equiv K_\perp / \tilde K^0$.
Thus $R_S$ and $R_L$ have simple geometric interpretation 
as the 'side' and 'long' widths of the distribution function,
while $R_O$ is a mixture of the 'out' width, time spread, {\em and}
the $x_O-\tilde t$ correlation.

{\em Covariant parton transport theory.}
We consider here, as in Refs. 
\cite{ZPC,nonequil,v2},
the simplest but nonlinear
form of Lorentz-covariant Boltzmann transport theory
in which the on-shell phase space density $f(x,\vp)$,
evolves with an elastic $2\to 2$ rate as
\bea
p_1^\mu \partial_\mu \tilde f_1 &=& \tilde S(x, \vp_1) 
+ \frac{\pi^4}{2} \int\limits_2\!\!\!\!
\int\limits_3\!\!\!\!
\int\limits_4\!\!
\left(
\tilde f_3 \tilde f_4 - \tilde f_1 \tilde f_2
\right)
\left|{\overline{\cal M}}_{12\to 34}\right|^2 \nonumber\\
&&\qquad\qquad\qquad\quad\times \  \delta^4(p_1{+}p_2{-}p_3{-}p_4)
 \ .
\label{Eq:Boltzmann_22}
\eea
Here $|\overline {\cal M}|^2$ is the polarization averaged scattering matrix element
squared,
the integrals are shorthands
for $\int_i \equiv \int d^3 p_i / [(2\pi)^3 E_i]$,
while $\tilde f_j \equiv (2\pi)^3 f(x, \vp_j)$.
The initial conditions are specified by the source function $\tilde S(x,\vp)$.
For our applications below,
we  neglect quark degrees of freedom and
interpret  $f(x,\vp)$ as describing
an ultrarelativistic massless gluon gas (8 colors, 2 helicities).

Eq. (\ref{Eq:Boltzmann_22}) can be  extended
to include inelastic matrix elements,
such as $gg \leftrightarrow ggg$,
and proper Bose or Fermi statistics can be introduced as well.
However, at present there is no {\em practical} algorithm
to compute accurate numerical solutions to such transport equations
on the workstations available to us.
Therefore, the present study is limited to the classical case with elastic $2\to 2$ interactions.

The elastic gluon scattering matrix elements in dense parton systems
were modeled with the isotropic form $ d\sigma_{el}/dt = \sigma_0 (s) / s$,
as justified by our previous study\cite{v2}.
We showed in \cite{v2} that the covariant transport solutions
do not depend {\em explicitly}
on the differential cross section
but only on the {\em transport opacity}
\be
\chi\equiv \frac{\sigma_{tr}}{\sigma_{el}} \langle n \rangle
\approx \sigma_{tr}
\langle \int dz  
\rho\left({\bf x}_0+ z\hat{\bf n},\tau=\frac{z}{c} \right)\rangle
\;\; ,
\label{Eq:tropacity}
\ee
where $\sigma_{tr}(s) \equiv \int d\sigma_{el} \sin^2\theta_{cm}$
is the transport cross section (in our case, $\sigma_{tr} = 2\sigma_0/3$),
and $\langle n \rangle$
is the average number of scatterings per  parton.
For a fixed nuclear geometry,
a given transport opacity $\chi$ represents
a {\em whole class} of initial conditions and partonic matrix elements,
as demonstrated by the approximate proportionality\cite{v2}
$\chi\propto \sigma_{tr} dN_g(\tau_0)/d\eta$.

We solved Eq. (\ref{Eq:Boltzmann_22}) numerically
via the MPC parton cascade algorithm\cite{MPCsource}.
MPC utilizes 
the particle subdivision technique\cite{ZPC},
which is essential to eliminate numerical artifacts caused by
frame-dependent collision ordering and
acausal (superluminal) propagation due to
action at a distance\cite{nonequil}.
For initial partonic densities expected at RHIC,
the severe violation of Lorentz covariance
in the naive cascade algorithm that employs no subdivision
artificially reduces elliptic flow and heats up the
$p_\perp$ spectra\cite{v2,finrange}.

We modeled central Au+Au collisions at RHIC 
with the minijet initial conditions used in Ref. \cite{v2}.
The evolution started from a 
longitudinally boost invariant Bjorken tube 
at proper time $\tau_0=0.1$~fm/$c$,
with locally isotropic momentum distribution 
and uniform  pseudorapidity $\eta\equiv 1/2 \log[(t+z)/(t-z)]$ distribution
between $|\eta| < 5$ with $dN_g(\tau_0)/d\eta=1050$.
The initial transverse density distribution was
proportional to the binary collision distribution for 
two Woods-Saxon distributions,
while the $p_\perp$ distribution was a thermal fit with $T=700$ MeV as in
Ref. \cite{v2}.

{\em Pion freezeout results.}
The freezeout distribution $d^4N/d^4 x$ was defined
as the distribution of space-time coordinates for 
 the last interaction point of the test particles.
Our strong simplifying assumption is that this point is not affected by 
hadronization.
We also neglected resonance contributions to the pion yield.
The same $1g\to 1\pi$ hadronization model was applied as in Ref. \cite{v2}
motivated by parton-hadron duality. 

\begin{figure}[hbpt] 
\hspace*{-0.2cm}\epsfig{file=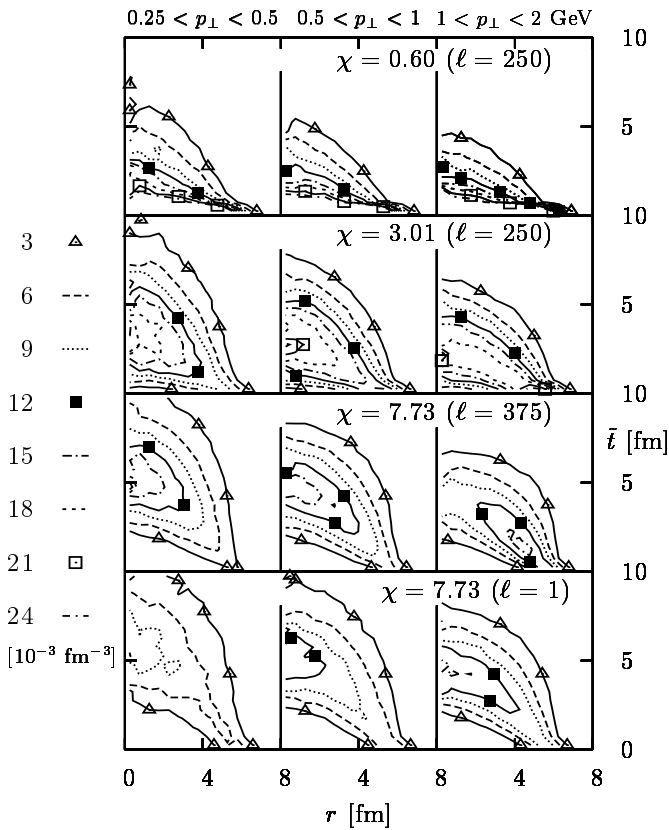,height=4.4in,width=3.5in,clip=5,angle=0} 
\begin{minipage}[t]{8.6cm}  
\vspace*{-0.5cm} 
\caption{\label{fig:rtau} \label{fig:rtau_l} Contour lines for freezeout distributions $dN/rdrd\tilde t$ from MPC for Au+Au at RHIC 
as a function of $p_\perp$ and transport opacity.
Distributions are normalized to unity in each $p_\perp$ bin.}
\end{minipage} 
\end{figure} 
Fig.~\ref{fig:rtau} shows the pion freezeout distribution $d^2N/r dr 
d\tilde t$ (where $r\equiv \sqrt{x_O^2 + x_S^2}$). 
Three ranges of the average pair transverse momentum
are considered. The variation of the distributions with transport opacities 
$\chi = 0.60$, $3.01$, and $7.73$ is shown.
For $dN_g/d\eta = 1050$ these correspond to $\sigma_{el} = 0.6$, $3$,
and $7.5$ mb.
Unlike the sharp freezeout imposed in hydrodynamical models,
the transport theory freezeout is a continuous, {\em evaporation-like}
 process\cite{nonequil}.
For a given (nonzero) opacity,
the larger the $p_\perp$ of the particle,
the earlier it decouples and the closer it is to the surface of the nuclei.
Low-$p_\perp$ particles freeze out from the center at late times,
while high-$p_\perp$ ones escape from the surface early.
Furthermore, 
a larger opacity increases decoupling times, especially for low-$p_\perp$ 
particles.

Fig.~\ref{fig:rtau_l} also demonstrates the importance of using a 
covariant algorithm, such as MPC, for solving Eq. (\ref{Eq:Boltzmann_22}).
For $\chi = 7.73$,
the covariant freezeout $dN/rdrd\tilde t$ distribution (third row)
differs significantly from that obtained with the naive noncovariant cascade
method, i.e., without particle subdivision (fourth row).
At the center, the covariant freezeout density decreases with increasing $p_\perp$, opposite to the monotonic increase shown by the noncovariant result.
Moreover,
for the naive algorithm,
the freezeout density peaks at the center ($r\approx 0$) for
all $p_\perp$ bins shown,
while for the covariant result the maximum
moves out towards the surface as $p_\perp$ increases.

\begin{figure*}[hbpt]
\hspace*{-0.3cm }\epsfig{file=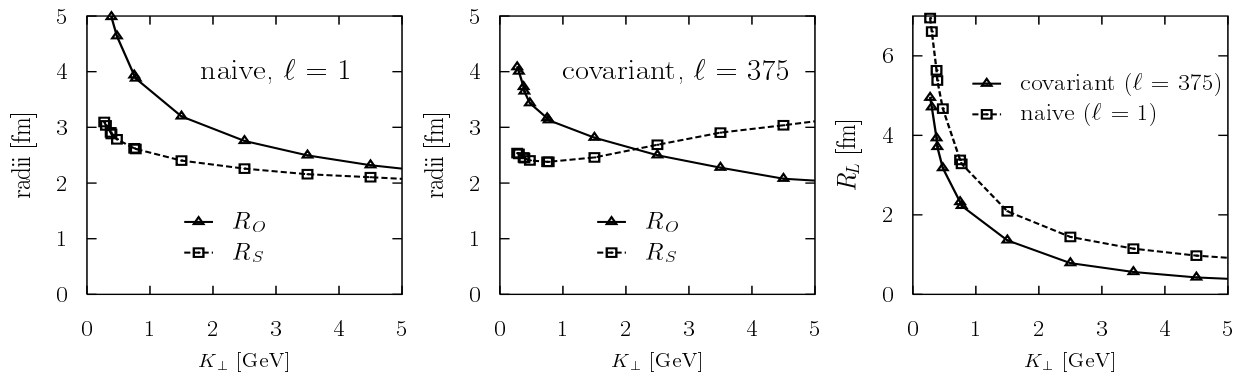,height=2.in,width=6.5in,clip=5,angle=0} 
\begin{minipage}[t]{8.6cm}  
\vspace*{-0.5cm } 
\caption{\label{fig:hbt_l} Strong parton subdivision dependence of
the pion HBT radii as a function of $K_\perp$ 
for transport opacity $\chi = 7.73$.
}
\end{minipage}
\end{figure*}
Naturally, Lorentz-violating artifacts also affect the geometric source radii
as shown in Fig.~\ref{fig:hbt_l}
for $\chi = 7.73$. 
Compared to the naive cascade algorithm,
the covariant one gives smaller $R_O$ and $R_L$
for all $K_\perp$.
The discrepancy is larger for $R_L$,
and for both radii increases as $K_\perp$ decreases,
leading to about 1 and 2 fm corrections, respectively, at $K_\perp \approx 0.2$
GeV.
It is remarkable that for $R_S$ the correction changes
sign as $K_\perp$ increases,
and that consequently the naive cascade method
yields $R_O / R_S > 1$ for all $K_\perp$,
while the covariant result shows $R_O < R_S$ for $K_\perp > 2$ GeV.
In the transport $R_O < R_S$ occurs because of
strong positive dynamical $x_O-\tilde t$ correlations.

\begin{figure*}[hbpt] 
\hspace*{-0.2cm }\epsfig{file=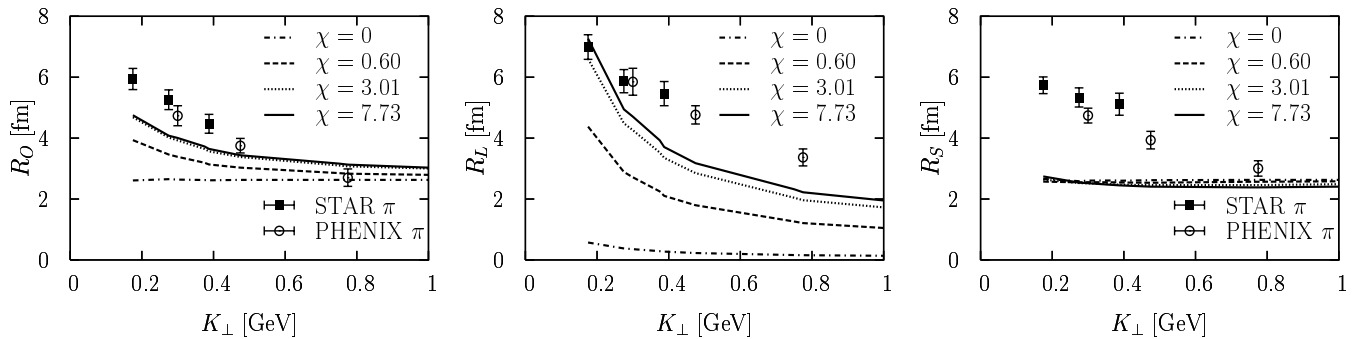,height=1.75in,width=7in,clip=5,angle=0}
\begin{minipage}[t]{8.6cm}  
\vspace*{-0.5cm} 
\caption{\label{fig:hbt} HBT radii as a function of $K_\perp$ and transport opacity. Data from Refs. \cite{STARhbt,PHENIXhbt} are shown.
}
\end{minipage}
\end{figure*} 
Fig.~\ref{fig:hbt} shows a comparison to the
HBT radii measured at RHIC.
In the transverse opacity range $\chi \sim 0 - 8$ we studied,
the classical transport results are smaller than the observed $R_O$ and $R_L$.
This is in sharp contrast to ideal hydrodynamics,
which overpredicts both radii\cite{hydroHBT}.
The monotonic dependence of $R_O$ and $R_L$ on transport opacity
suggests that better agreement with data may be possible
with larger opacity $\chi ~ \sim 20-30$, which unfortunately are numerically
impractical as yet. The need for 
such high opacities is also suggested 
by the elliptic flow $v_2(p_\perp)$ saturation analysis\cite{v2}.

In Fig.~\ref{fig:hbt},
$R_L$ is obviously small for zero opacity
because freezeout occurs then at the formation time $\tau = \tau_0$.
Note that $R_L^2 \approx \tau^2 [\Delta(\eta - y)]^2$
depends on the decoupling time and the strength of the $\eta-y$
correlation.
For our thermally correlated initial condition
$[\Delta(\eta - y)]^2\approx T/m_\perp$ at $\tau_0 = 0.1$ fm$/c$.

However,
as the transport opacity increases,
$R_L$ grows rapidly 
because the decoupling time increases as is evident from Fig.~\ref{fig:rtau}.
The largest increase $\tau/\tau_0 \sim R/\tau_0 \sim 50$
is for low-$p_\perp$ partons, which freeze out latest.
The observed $R_L(K_\perp)$ is a sensitive
probe of the product of the freezeout proper time and $\Delta(\eta-y)$.
Thus a perfect inside-outside correlation, i.e., $\eta = y$,
as assumed in classical Yang-Mills approaches,
cannot be reconciled with
the RHIC $R_L$ data, without final state interactions.

In our approach, the main remaining 
 puzzle in Fig.~\ref{fig:hbt}
is the predicted $R_S(K_\perp) \approx const \approx 3$ fm
that is peculiarly  independent of the transport opacity
and underestimates significantly the observed side radius.
This suggests that $R_S$ is insensitive to
the early partonic collective dynamics.
The same underestimate of $R_S$ 
has been found in hydrodynamical calculations as well\cite{hydroHBT}.
The $R_S$ problem may be related to
the assumed longitudinally boost invariant dynamics
in both approaches. However, it also could be related to our neglect of 
hadronic resonances treated in \cite{AdrianHBT,Lin:2002gc}.
We also note that for more spherically symmetric initial condition,
even ideal hydrodynamical solutions\cite{MiklosBP} 
 exhibit $R_O/R_S \sim 1$ with larger side radii.

{\em Conclusions.}
Using the MPC technique,
we investigated the effect of early phase dissipative partonic dynamics 
on the decoupling geometry in heavy-ion collisions in the RHIC energy domain.
The pion freezeout distribution at midrapidity was found to be 
sensitive to the transport opacity of partons as in \cite{Lin:2002gc}.
The transport freezeout process is similar to evaporation:
high-$p_\perp$ particles freeze out early from the surface,
while low-$p_\perp$ ones decouple late from the center.
For $K_\perp \lesssim 2$ GeV, $R_O$ was found to become smaller than 
$R_S$ indicating that  positive $x_O-\tilde t$ dynamical correlations
are strongly sensitive to finite mean free path effects.

We also demonstrated that the naive cascade
algorithms without high particle subdivision lead to 
large numerical artifacts in the freezeout distribution
due to violation of Lorentz covariance.
These artifacts enhance the out and long radii,
especially at low $K_\perp$,
while increase(reduce) $R_S$ for $K_\perp$ below(above) $\approx 2$ GeV.
The MPC technique makes it possible to avoid those artifacts.

While we showed that a decreasing $R_{out}/R_{side}<1$ 
with $K_\perp$ can arise from covariant parton transport
dynamics, the momentum scale where this happens is not realistic
due to the simplified local $g\rightarrow \pi$ hadronization scheme 
employed as well as the neglect of hadronic transport.
A consistent explanation of all the differential features
of HBT correlations will have to include in the future a more realistic 
covariant model of hadronization as well as 
maintain covariance during the hadronic final state interactions.

{\em Acknowledgments:}  This work is  supported by the Director, 
Office of Science, Office of High Energy and Nuclear Physics, 
Division of Nuclear Physics, of the U.S. Department of Energy 
under Grants No. DE-FG02-93ER40764 and DE-FG02-01ER41190.
We acknowledge the Parallel Distributed Systems Facility
at the National Energy Research Scientific Computing Center
for providing computing resources.

\end{document}